\documentclass[10pt]{article}

\usepackage{graphics}
\usepackage{graphicx}

\usepackage[a4paper, left=35mm,right=35mm,top=34mm,bottom=34mm]{geometry}
\usepackage[utf8]{inputenc}
\usepackage[T1]{fontenc}
\usepackage[english]{babel}

\usepackage{enumerate}
\usepackage{graphicx}
\usepackage{hyperref}
\hypersetup{
    colorlinks=true,
    linkcolor=blue,
    filecolor=magenta,      
    urlcolor=cyan,
}
\usepackage{listings}
\usepackage{color}

\definecolor{dkgreen}{rgb}{0,0.6,0}
\definecolor{gray}{rgb}{0.5,0.5,0.5}
\definecolor{mauve}{rgb}{0.58,0,0.82}
\lstdefinelanguage{MRGC++}{%
  language=C++,
  morekeywords={T, U, MPI_Irecv, MPI_Isend, MPI_Allreduce, MPI_Waitall, Compute, Map, abs, max, Swap, MPI_Recv_init, MPI_Send_init, MPI_Startall, Copy, Init, InitRecv, InitSend, InitAllReduce, Send, Recv, AllReduce, Finalize, InitSnapshot, Snapshot, SwitchAsync, SnapReduce, MPI_Test, MPI_Start}
}
\usepackage{mathtools,amsthm,amssymb,amsfonts}
\usepackage{algorithm}
\usepackage{algorithmic}
\makeatother
\theoremstyle{plain}

\theoremstyle{definition}

\theoremstyle{remark}

\usepackage{caption} 
\usepackage{fancyhdr}

\author{
  {\normalsize Guillaume Gbikpi-Benissan}\thanks{Peoples' Friendship University of Russia (RUDN), Russian Federation
    (correspondence, guibenissan@gmail.com).}
  \and
  {\normalsize Fr\'ed\'eric Magoul\`es}\thanks{CentraleSup\'elec, Universit\'e Paris-Saclay, France.}
}
\title{Distributed asynchronous convergence detection without detection protocol}
\date{}

\begin{document}
\maketitle
\thispagestyle{fancy}

\begin{abstract}
\noindent In this paper, we address the problem of detecting the moment when an ongoing asynchronous parallel iterative process can be terminated to provide a sufficiently precise solution to a fixed-point problem being solved. Formulating the detection problem as a global solution identification problem, we analyze the snapshot-based approach, which is the only one that allows for exact global residual error computation. From a recently developed approximate snapshot protocol providing a reliable global residual error, we experimentally investigate here, as well, the reliability of a global residual error computed without any prior particular detection mechanism. Results on a single-site supercomputer successfully show that such high-performance computing platforms possibly provide computational environments stable enough to allow for simply resorting to non-blocking reduction operations for computing reliable global residual errors, which provides noticeable time saving, at both implementation and execution levels.
\end{abstract}


\section{Introduction}

One of the major questions which arise when implementing asynchronous iterations in distributed solvers consists of finding a mechanism to detect when convergence is reached, so that one can terminate the solving process and deliver an approximated solution actually satisfying a desired precision. On efficiency aspects, centralized detection protocols obviously suffer from scaling limits on very large distributed platforms, and more elaborated mechanisms have to face possible termination delays. On the other hand, effective convergence is hardly always guaranteed when resorting to assumptions-based light protocols. One thus has to figure out what is the most appropriate choice according to his particular parallel configuration.

To be more precise, let a sequence $\{x^{k}\}_{k \in \mathbb{N}}$ of vectors be generated by an asynchronous parallel iterative procedure, in order to find the solution $x^{*}$ of a fixed-point problem
\[
x - f(x) = 0.
\]
Let $r$ be a residual error evaluation function such that
\[
r(x^{k}) \simeq 0 \quad \implies \quad x^{k} \simeq x^{*}, \qquad k \in \mathbb{N}.
\]
In the asynchronous iterations context,
$\{x^{k}\}_{k \in \mathbb{N}}$
is actually implicit, and one only explicitly handles
$p$
parallel
sequences
$\{x_{1}^{k^{(1)}}\}_{k^{(1)} \in \mathbb{N}}$ to $\{x_{p}^{k^{(p)}}\}_{k^{(p)} \in \mathbb{N}}$,
of local subvectors which can be gathered to provide global potential solution vectors.
The asynchronous convergence detection problem therefore consists of being able to determine, in a non-blocking way, and as quickly as possible, the moment when
\[
r(\bar x) \simeq 0, \qquad\qquad \bar x =
\begin{bmatrix}
x_{1}^{k^{(1)}} & \cdots & x_{p}^{k^{(p)}}
\end{bmatrix}^{\mathsf T},
\qquad k^{(1)}, \ldots, k^{(p)} \in \mathbb{N}.
\]

The main distributed approaches are based on
\begin{itemize}
\item modification of the iterative procedure to ensure finite-time termination (see, e.g., \cite{BertTsit1989b, ElBaz1996, SavBert1996}),
\item explicit evaluation of the residual error $r(\bar x)$ from global state snapshot (see \cite{SavBert1996, MagGBen2018}), implemented in the asynchronous iterations programming library JACK2~\cite{MagGBen2018b},
\item predictive approximation of the number of iterations required to reach convergence (see \cite{EvChik1998}),
\item monitoring of both consistency and persistence of local convergence (see \cite{BahiEtAl2005, BahiEtAl2008}), implemented in the asynchronous iterations programming libraries JACE~\cite{BahiEtAl2004}, CRAC~\cite{CoutDom2007}, JACK~\cite{MagGBen2017} and JACK2~\cite{MagGBen2018b},
\item evaluation of diameter of solutions nested sets by means of performing ``macro-iterations'' (see \cite{MielEtAl2008}).
\end{itemize}

Modifying the iterative procedure has got the major drawback of being intrusive and even requiring additional assumptions over the asynchronous iterative model. The monitoring-based and the prediction-based approaches can lead to untimely termination, and would therefore require a blocking check after convergence detection. The monitoring-based approach could be made fully reliable by applying it jointly with the approach making use of nested sets, which was mainly focusing on mathematical correctness. Still, intrusive piggybacking techniques remain needed at implementation level, to define different types of computation messages. While successfully being completely non-intrusive, the snapshot approach proposed in \cite{SavBert1996} however requires to introduce computation data into snapshot messages, which leads to a $\mathcal O(n)$ communication overhead. In \cite{MagGBen2018}, a $\mathcal O(1)$ snapshot message size is achieved, but at the cost of assuming a known bound on communication delays.

The analysis in \cite{MagGBen2018} shows the evaluation of an approximated residual error while explicitly bounding its difference with the exact one. Roughly, it allows for a non strictly consistent snapshot from which the residual error is then evaluated through a non-blocking reduction operation. We are therefore investigating, in this paper, to which extent such a snapshot could be non consistent, which even allows us to consider no control at all, meaning not performing any prior snapshot protocol.

Section~\ref{sec:bg} recalls the asynchronous convergence detection problem. In Section~\ref{sec:main}, we present snapshot-based residual error evaluation and how it leads to consider no detection protocol at all. Experimental investigation is conducted in Section~\ref{sec:ex} and our conclusions follow in Section~\ref{sec:con}.

\section{Computational background}
\label{sec:bg}

\subsection{Asynchronous iterations}

Let
\[
x - f(x) = 0
\]
be a fixed-point problem admitting a unique solution $x^{*}$. The classical fixed-point iterative method for finding $x^{*}$ consists of generating a sequence $\{x^{k}\}_{k \in \mathbb{N}}$ such that
\begin{equation}
\label{eq:sia}
x^{k+1} = f(x^{k}),
\end{equation}
and that, for any given $x^{0}$,
\[
\lim_{k \to +\infty} x^{k} = x^{*}.
\]
In a parallel context with $p$ processing units, let us consider a decomposition of the form
\[
x =
\begin{bmatrix}
x_{1} & \cdots & x_{p}
\end{bmatrix}^{\mathsf T},
\qquad
f(x) =
\begin{bmatrix}
f_{1}(x) & \cdots & f_{p}(x)
\end{bmatrix}^{\mathsf T},
\]
which leads to a parallel iterative method given by
\[
x_{i}^{k+1} = f_{i}(x_{1}^{k}, \ldots, x_{p}^{k}), \qquad \forall i \in \{1, \ldots, p\}.
\]
While each iteration \eqref{eq:sia} is therefore made parallel, there is still one global sequence of iterations, which leads to possible idle times for interprocess communication, especially in distributed-memory environments requiring message passing. Figure~\ref{fig:spia} shows message exchange in the following possible example of execution with $p = 2$:
\begin{figure}
\begin{center}
\includegraphics[scale=0.55]{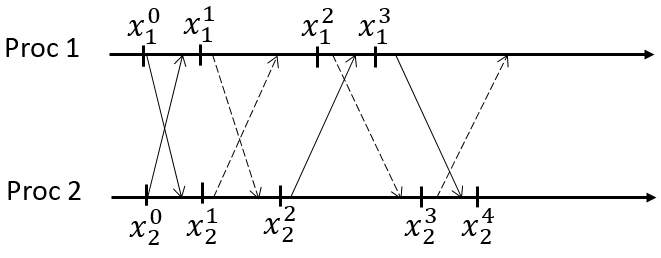}
\caption{Message exchange in a classical parallel iterative process, where dash arrows indicate induced idle time.}
\label{fig:spia}
\end{center}
\end{figure}
\[
\begin{array}{ccc}
x_{1}^{1} := f_{1}(x_{1}^{0}, x_{2}^{0}) &\qquad\qquad\qquad& x_{2}^{1} := f_{2}(x_{1}^{0}, x_{2}^{0})\\
\text{idle} &\qquad\qquad\qquad& \text{idle}\\
x_{1}^{2} := f_{1}(x_{1}^{1}, x_{2}^{1}) &\qquad\qquad\qquad& x_{2}^{2} := f_{2}(x_{1}^{1}, x_{2}^{1})\\
x_{1}^{3} := f_{1}(x_{1}^{2}, x_{2}^{2}) &\qquad\qquad\qquad& \text{idle}\\
\text{idle} &\qquad\qquad\qquad& x_{2}^{3} := f_{2}(x_{1}^{2}, x_{2}^{2})\\
\text{idle} &\qquad\qquad\qquad& x_{2}^{4} := f_{2}(x_{1}^{3}, x_{2}^{3})
\end{array}
\]

Having arisen with a first rigorous analysis by \cite{ChazMir1969}, asynchronous iterations consist of considering actual parallel sequences of iterations where each processing unit can compute and exchange data at its own pace. Communication is therefore completely overlapped, as shown in Figure~\ref{fig:apia} where the message exchange pattern corresponds to the following example of execution with $p = 2$:
\begin{figure}
\begin{center}
\includegraphics[scale=0.55]{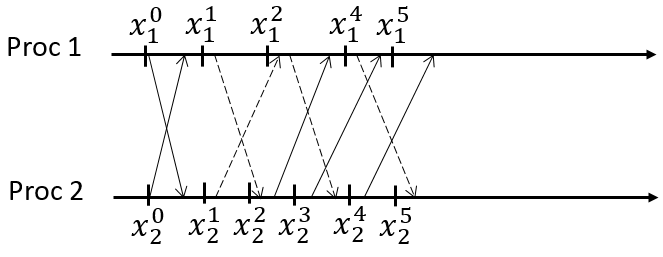}
\caption{Message exchange in an asynchronous parallel iterative process, where dash arrows indicate delays.}
\label{fig:apia}
\end{center}
\end{figure}
\[
\begin{array}{lll}
x_{1}^{1} := f_{1}(x_{1}^{0}, x_{2}^{0}) &\qquad\qquad\qquad& x_{2}^{1} := f_{2}(x_{1}^{0}, x_{2}^{0})\\
x_{1}^{2} := f_{1}(x_{1}^{1}, x_{2}^{0}) &\qquad\qquad\qquad& x_{2}^{2} := f_{2}(x_{1}^{0}, x_{2}^{1})\\
x_{1}^{3} := x_{1}^{2} \ \text{(implicit)} &\qquad\qquad\qquad& x_{2}^{3} := f_{2}(x_{1}^{1}, x_{2}^{2})\\
x_{1}^{4} := f_{1}(x_{1}^{3}, x_{2}^{2}) &\qquad\qquad\qquad& x_{2}^{4} := f_{2}(x_{1}^{2}, x_{2}^{3})\\
x_{1}^{5} := f_{1}(x_{1}^{4}, x_{2}^{3}) &\qquad\qquad\qquad& x_{2}^{5} := f_{2}(x_{1}^{2}, x_{2}^{4})
\end{array}
\]
We thus see that a global sequence $\{x^{k}\}_{k \in \mathbb{N}}$ can still be modeled, if one consider both implicit operations and delayed data in the sense that $x_{i}^{k+1}$, $i \in \{1, \ldots, p\}$, is no more necessarily computed on the basis of $x_{j}^{k}$, $j \in \{1, \ldots, p\}$, but instead using some possibly delayed $x_{j}^{\tau_{j}^{i}(k)}$ with
\[
\tau_{j}^{i}(k) \le k.
\]
This leads to a more general iterative model given by
\begin{equation}
\label{eq:apia}
x_{i}^{k+1} = \left \{
\begin{array}{ll}
f_{i}\left(x_{1}^{\tau_{1}^{i}(k)}, \ldots, x_{p}^{\tau_{p}^{i}(k)}\right), & \forall i \in P^{(k)},\\
x_{i}^{k}, & \forall i \notin P^{(k)},
\end{array}
\right.
\end{equation}
where $P^{(k)}$ gives the set of components being explicitly updated at iteration $k$, i.e.,
\[
P^{(k)} \subseteq \{1, \ldots, p\}.
\]
While, therefore, both computation and communication patterns can be completely random, it is still required that any component $x_{i}$ keeps being eventually updated, i.e.
\[
\forall i \in \{1, \dots, p\}, \quad \operatorname{card}\{k \in \mathbb{N} \ | \ i \in P^{(k)}\} = +\infty,
\]
and that any component $x_{j}$ used for update also keeps being eventually updated, i.e.,
\begin{equation}
\label{eq:ai_ass}
\forall i, j \in \{1, \dots, p\}, \quad \lim_{k \to +\infty} \tau_{j}^{i}(k) = +\infty.
\end{equation}
We refer to \cite{BertTsit1989} for a complete introduction to the asynchronous iterations theory, including convergence conditions of such algorithms.

\subsection{Convergence detection}

Let $S^{*}$ be a set of admissible solutions, according to a desired precision $\varepsilon \in \mathbb{R}$. Let then $r$ be a residual error evaluation function, i.e.,
\[
r(x) < \varepsilon \quad \implies \quad x \in S^{*}, \qquad k \in \mathbb{N}.
\]
Such a function is usually distributed as
\[
r(x) = \sigma(r_{1}(x), \ldots, r_{p}(x)),
\]
where $\sigma$ is a reduction function. If, for instance, $r$ is given by
\begin{equation}
\label{eq:res}
r(x) = \|x - f(x)\|_{2},
\end{equation}
where $\|.\|_{2}$ denotes the Euclidean norm, then we can have
\[
r_{i}(x) = \left(\|x_{i} - f_{i}(x)\|^{(i)}\right)^{2}, \qquad \sigma(\alpha_{1}, \ldots, \alpha_{p}) = \left(\sum_{j=1}^{p} \alpha_{j}\right)^{1/2},
\]
where $\|.\|^{(i)}$ denotes a norm defined on the vector space wherein the component $x_{i}$ lies.

Nevertheless, it follows from the asynchronous iterative model \eqref{eq:apia} that the generated global sequence $\{x^{k}\}_{k \in \mathbb{N}}$ is implicit. In practice, one actually handles $p$ local sequences $\{x_{1}^{k^{(1)}}\}_{k^{(1)} \in \mathbb{N}}$ to $\{x_{p}^{k^{(p)}}\}_{k^{(p)} \in \mathbb{N}}$, while not necessarily having
\begin{equation}
\label{eq:distsync}
x_{i}^{k^{(i)}} = x_{i}^{k} \quad \iff \quad k^{(i)} = k, \qquad i \in \{1, \ldots, p\},
\end{equation}
which is satisfied in classical synchronous case given by
\[
\tau_{j}^{i}(k) = k, \quad P^{(k)} = \{1, \ldots, p\}, \qquad \forall i,j \in \{1, \ldots, p\}, \quad \forall k \in \mathbb{N}.
\]
There is therefore no trivial way to evaluate
\begin{equation}
\label{eq:convdetect}
r(\bar x) < \varepsilon, \qquad \bar x =
\begin{bmatrix}
x_{1}^{k^{(1)}} & \cdots & x_{p}^{k^{(p)}}
\end{bmatrix}^{\mathsf T},
\qquad\qquad k^{(1)}, \ldots, k^{(p)} \in \mathbb{N},
\end{equation}
as $\bar x$ is not directly accessible, due to Equation~\eqref{eq:distsync} not being satisfied. Main attempts to solve the convergence detection problem thus resulted in indirect ways of inducing Equation~\eqref{eq:convdetect}, either accurately or approximately, but requiring more or less intrusive implementation, in the sense that the detection mechanism cannot be completely encapsulated out of the computation part of the code, just like in classical synchronous case (see, e.g., \cite{BertTsit1989b, ElBaz1996, MielEtAl2008, BahiEtAl2008}). The first distributed algorithm for explicitly evaluating Equation~\eqref{eq:convdetect} goes back to \cite{SavBert1996}, where a simple effective snapshot approach is proposed to build $\bar x$. The whole snapshot and subsequent residual error evaluation operations have then been recently successfully implemented in a non-intrusive way, within the library JACK2~\cite{MagGBen2018b}, for any norm
\[
\|x\|_{l} = \left(\sum_{i=1}^{p}(\|x_{i}\|_{l})^{l}\right)^{1/l}, \qquad l \ge 1,
\]
including the maximum norm
\[
\|x\|_{\infty} = \max_{i=1}^{p}\|x_{i}\|_{\infty}.
\]

\section{Computing asynchronous residual error}
\label{sec:main}

\subsection{Exact residual error}

Computing an exact residual error requires to build a global vector $\bar x$ and then to evaluate $r(\bar x)$, all of this in a distributed non-blocking way, concurrently to the asynchronous iterative process. Evaluating $r(\bar x)$ is a classical reduction operation which only needs to be non-blocking. Non-blocking reduction operations are provided by JACK2~\cite{MagGBen2018b} and now, also by the third version of the Message Passing Interface (MPI) specification.

\begin{figure}
\begin{center}
\includegraphics[scale=0.55]{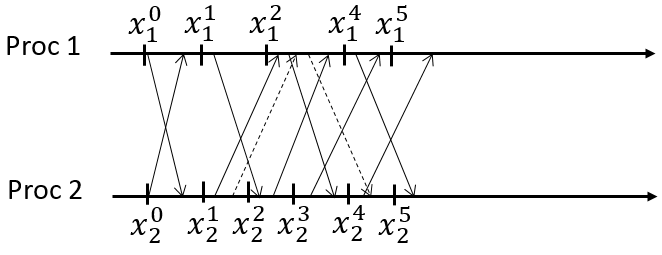}
\caption{Snapshot message exchange (dash arrows) in an asynchronous parallel iterative process.}
\label{fig:apia_cls}
\end{center}
\end{figure}

To build $\bar x$ while being in a first-in-first-out (FIFO) communication environment, where messages on a unidirectional communication link are delivered in the order in which they have been emitted, the general snapshot algorithm from \cite{ChanLamp1985} has been recently adapted for asynchronous convergence detection by \cite{MagGBen2018}. The corresponding event-based protocol is given as follows:
\begin{itemize}
\item \textbf{On} local convergence \textbf{or} first reception of snapshot message \textbf{do}
\begin{enumerate}
\item Record current local component $x_{i}^{(i),k^{(i)}}$ as $\bar x_{i}^{(i)}$
\item Send snapshot message on all outgoing communication links
\end{enumerate}
\item \textbf{On} reception of snapshot message \textbf{do}
\begin{enumerate}
\item Record last dependence $x_{j}^{(i),k^{(i)}}$ received on the corresponding incoming communication link as $\bar x_{j}^{(i)}$
\end{enumerate}
\end{itemize}
To fully express the local independent state of each subprocess $i \in \{1, \ldots, p\}$, we add the subscript $(i)$ to components $x_{j}$, $j \in \{1, \ldots, p\}$, to denote the specific version of data accessed by the subprocess $i$. Local convergence here, on subprocess $i$, is usually given by
\[
r_{i}(x^{(i),k^{(i)}}) < \varepsilon,
\]
for instance. The following example of execution, still using implicit global iterations, can correspond to the snapshot message exchange illustrated in Figure~\ref{fig:apia_cls}. Recorded local components are on the left and recorded dependencies, on the right of each subprocess.
\[
\begin{array}{lllllllllll}
&&x_{1}^{1} := f_{1}(x_{1}^{0}, x_{2}^{0}) &&&\qquad\qquad&x_{2}^{1}&\quad& x_{2}^{1} := f_{2}(x_{1}^{0}, x_{2}^{0})&&\\
&&x_{1}^{2} := f_{1}(x_{1}^{1}, x_{2}^{0}) &&&\qquad\qquad&&& x_{2}^{2} := f_{2}(x_{1}^{0}, x_{2}^{1})&&\\
x_{1}^{2}&\quad&x_{1}^{3} := x_{1}^{2}&\quad&x_{2}^{1}&\qquad\qquad&&& x_{2}^{3} := f_{2}(x_{1}^{1}, x_{2}^{2})&&\\
&&x_{1}^{4} := f_{1}(x_{1}^{3}, x_{2}^{2}) &&&\qquad\qquad&&& x_{2}^{4} := f_{2}(x_{1}^{2}, x_{2}^{3})&\quad&x_{1}^{2}\\
&&x_{1}^{5} := f_{1}(x_{1}^{4}, x_{2}^{3}) &&&\qquad\qquad&&& x_{2}^{5} := f_{2}(x_{1}^{2}, x_{2}^{4})&&
\end{array}
\]
One therefore rightly obtains
\[
\bar x = \bar x^{(1)} = \bar x^{(2)} =
\begin{bmatrix}
x_{1}^{2} & x_{2}^{1}
\end{bmatrix}^{\mathsf T}.
\]

The FIFO requirement means that a snapshot message sent after a computation message on the same unidirectional communication link cannot be delivered before the computation message. This guarantees the consistency of the snapshot, which can be easily checked on the preceding example of execution. However, the MPI specification does not ensure FIFO delivery for messages of different types, which makes it hard to implement such a protocol in a non-intrusive way within MPI environments. To overcome this limit, \cite{SavBert1996} suggested to include computation data into snapshot messages, which leads to the following protocol:
\begin{itemize}
\item \textbf{On} local convergence \textbf{do}
\begin{enumerate}
\item Record current local component $x_{i}^{(i),k^{(i)}}$ as $\bar x_{i}^{(i)}$
\item Send snapshot message $x_{i}^{(i),k^{(i)}}$ on all outgoing communication links
\end{enumerate}
\item \textbf{On} reception of snapshot message $x_{j}^{(j),k^{(j)}}$ \textbf{do}
\begin{enumerate}
\item Record received dependence $x_{j}^{(j),k^{(j)}}$ as $\bar x_{j}^{(i)}$
\end{enumerate}
\end{itemize}
We point out that in practice, the whole local component $x_{i}^{(i),k^{(i)}}$ does not need to be sent, but only the corresponding interface data needed by each neighbor subprocess $j$ for applying its local submapping $f_{j}$ (see, e.g., Fig.~4 in \cite{MagGBen2018b}, where only the content of the usual message sending buffers is sent as snapshot messages as well). Still, the size of the interface data can be significantly high, especially when the size of the problem increases while the number $p$ of subprocesses remains the same. To overcome this second limit, \cite{MagGBen2018} resorted to approximate residual error computation.

\subsection{Approximate residual error}

In \cite{MagGBen2018}, non-FIFO environments have been characterized by considering a degree of possibly out-of-order message delivery. Basically, the assumption is made that a message sent after $m$ other messages cannot be delivered before all of these $m$ preceding messages. Furthermore, considering practical situations where empty messages are more likely to be delivered faster than computation messages containing heavy interface data, one can also assume that a computation message sent after an empty snapshot message will always be delivered after this snapshot message. Such assumptions allowed \cite{MagGBen2018} to design the following non-FIFO asynchronous iterations snapshot protocol, which avoids including computation data into snapshot messages:
\begin{itemize}
\item \textbf{On} local convergence for $m$ successive iterations \textbf{do}
\begin{enumerate}
\item Record current local component $x_{i}^{(i),k^{(i)}}$ as $\bar x_{i}^{(i)}$
\item Send snapshot message on all outgoing communication links
\end{enumerate}
\item \textbf{On} reception of snapshot message \textbf{do}
\begin{enumerate}
\item Record last dependence $x_{j}^{(i),k^{(i)}}$ received on corresponding incoming communication link as $\bar x_{j}^{(i)}$
\end{enumerate}
\end{itemize}
Still, to cover cases where snapshot message delivery can be slower than one of computation message, a second type of snapshot message is sent after $m$ supplementary iterations, which confirms or discards the validity of the snapshot, depending on whether the local convergence state has remained persistent or not.

Computing an exact residual error in a non-FIFO environment is not possible with such a protocol since it potentially returns
\[
\bar x^{(1)} \ne \cdots \ne \bar x^{(p)}.
\]
However, let us define
\[
\widetilde r(x^{(1)}, \ldots, x^{(p)}) := \sigma(r_{1}(x^{(1)}), \ldots, r_{p}(x^{(p)})),
\]
so that we thus have
\[
\widetilde r(x, \ldots, x) = r(x).
\]
Then, the residual error computed from this approximate snapshot protocol is given by
\[
\widetilde r(\bar x^{(1)}, \ldots, \bar x^{(p)}).
\]
If, now, we take
\[
\bar x =
\begin{bmatrix}
\bar x^{(1)}_{1} & \cdots & \bar x^{(p)}_{p}
\end{bmatrix}^{\mathsf T},
\]
it has been shown in \cite{MagGBen2018} that the exact residual error is bounded as
\[
r(\bar x) < \widetilde r(\bar x^{(1)}, \ldots, \bar x^{(p)}) + c(p,m) \varepsilon,
\]
where $c(p,m)$ is a function of $p$ and $m$ which expression depends on that of $r$. This leads to
\[
\begin{aligned}
\widetilde r(\bar x^{(1)}, \ldots, \bar x^{(p)}) & < \varepsilon,\\
\widetilde r(\bar x^{(1)}, \ldots, \bar x^{(p)}) + c(p,m) \varepsilon & < (1 + c(p,m)) \varepsilon,\\
r(\bar x) & < (1 + c(p,m)) \varepsilon.
\end{aligned}
\]
If, therefore, convergence is given by
\[
r(\bar x) < \widetilde \varepsilon,
\]
one can ensure it by actually checking
\[
\widetilde r(\bar x^{(1)}, \ldots, \bar x^{(p)}) < \frac{\widetilde \varepsilon}{1 + c(p,m)}.
\]

The main idea behind the ability to rely on an approximate residual error comes from the required contraction property of fixed-point mappings. Typically, with a residual error given by Equation~\eqref{eq:res}, one can make use of
\[
\|f(x) - f(y)\| \le \alpha \|x-y\|, \qquad \alpha < 1,
\]
to also expect a bound on
\[
|\widetilde r(\bar x, \ldots, \bar x) - \widetilde r(\bar x^{(1)}, \ldots, \bar x^{(p)})|
\]
depending on differences
\[
\|\bar x - \bar x^{(i)}\|, \quad \forall i \in \{1, \ldots, p\}.
\]
Consistent snapshot ensures
\[
\bar x = \bar x^{(1)} = \cdots = \bar x^{(p)},
\]
which allows for exact residual error evaluation. The inconsistent snapshot protocol ensures a known bound
\begin{equation}
\label{eq:bound}
\|\bar x - \bar x^{(i)}\| < c^{(i)}(p,m) \varepsilon, \qquad i \in \{1, \ldots, p\},
\end{equation}
which allows for approximate residual error validation. This however depends on the knowledge of $m$. While having a bound on communication delays is necessary to satisfy the fundamental assumption \eqref{eq:ai_ass}, such a bound does not necessarily need to be known, and in this general case, $m$ could not be known either. This does not contradict the existence of a bound of the form of Equation~\eqref{eq:bound}, which, therefore is no longer dependent on the approximate snapshot protocol making use of a specific value of $m$. In this sense, by taking $\bar x^{(i)}$ in a completely arbitrary way, without any snapshot protocol, we can still set a residual error threshold $\varepsilon$ which would ensure convergence at a desired precision $\widetilde \varepsilon$. It would result that only successive non-blocking reduction operations have to performed until convergence is detected, just as in classical synchronous computation. We experimentally investigate such a protocol-free asynchronous convergence detection mechanism in the next section.

\section{Numerical experiments}
\label{sec:ex}

\subsection{Problem and experimental settings}

We experimented the protocol-free convergence detection approach by means of the JACK2~\cite{MagGBen2018b} library. The implementation of such a solution corresponds to Listing~4 in \cite{MagGBen2018b}, and both problem and experimental settings here are the same as in \cite{MagGBen2018b} as well. It basically consists of solving a convection-diffusion problem of the form
\[
\frac{\partial u}{\partial t} - \nu \Delta u + \vec a.\nabla u = s,
\]
on a cubic domain $[0, 1]^{3}$. Backward Euler and centered finite-differences schemes are used for time and space integration, resulting in successive sparse linear systems to be solved. The domain is partitioned into a grid in the $(x,y)$-plane, such that each subdomain contains the whole interval $[0, 1]$ in $z$-direction. The linear systems are solved by means of Jacobi relaxations at interface and Gauss-Seidel relaxations at interior nodes of the subdomains. Each processor core is assigned with one subdomain. The computational platform is an SGI ICE X supercomputer with FDR Infiniband network (56 Gb/s). Nodes are equipped with two 12-cores Intel Haswell Xeon CPUs at 2.30 GHz, 48 GB RAM, and SGI-MPT as MPI library.

Relating to the resolution of one time step linear system
\[
Ax = b, \qquad x \in \mathbb{R}^{n},
\]
where $n$ is the size of the problem, the following results will be reported:
\begin{itemize}
\item final residual errors
\[
r^{*} = \|A\widetilde x^{*} - b\|_{\infty},
\]
where $\widetilde x^{*}$ is the computed solution,
\item execution times in seconds, denoted as "wtime" (wall-clock time).
\item maximum numbers of iterations over the sets of MPI processes, denoted as $k_{\text{max}}$.
\end{itemize}

\subsection{Numerical results}

Our methodology is, first, to determine the stability of the computational platform, in order to set a residual error threshold $\varepsilon$ which will ensure
\begin{equation}
\label{eq:ex:precision}
r^{*} < \widetilde \varepsilon = 10^{-6}.
\end{equation}
To this end, we use a small-size version of the convection-diffusion problem, which does not require much computational time. Table~\ref{tab:stab_r} reports minimum and maximum final residual errors observed over several executions. It features the suggested protocol-free asynchronous iterations termination (PFAIT), the non-FIFO asynchronous iterations snapshot protocol 2 (NFAIS2~\cite{MagGBen2018}) which includes interface data into snapshot messages, and the NFAIS5~\cite{MagGBen2018} which corresponds to the approximate snapshot approach.
\begin{table}[!ht]
\begin{center}
{\small
\begin{tabular}{|c|c|c|c|}
\hline
& PFAIT & NFAIS2~\cite{MagGBen2018} & NFAIS5~\cite{MagGBen2018}\\
\hline
\begin{tabular}{c}
$p$\\
\hline
48\\
96\\
144\\
192\\
240\\
480\\
600
\end{tabular}
&
\begin{tabular}{c|c}
min $r^{*}$ & max $r^{*}$\\
\hline
1.28e-06 & 1.47e-06\\
8.52e-07 & 1.11e-06\\
9.55e-07 & 2.39e-06\\
1.03e-06 & 1.29e-06\\
9.28e-07 & 1.47e-06\\
9.69e-07 & 2.83e-06\\
9.39e-07 & 1.32e-06
\end{tabular}
&
\begin{tabular}{c|c}
min $r^{*}$ & max $r^{*}$\\
\hline
5.29e-07 & 6.81e-07\\
5.13e-07 & 6.33e-07\\
5.91e-07 & 6.05e-07\\
5.08e-07 & 5.83e-07\\
4.79e-07 & 5.55e-07\\
4.50e-07 & 5.76e-07\\
3.74e-07 & 5.28e-07
\end{tabular}
&
\begin{tabular}{c|c}
min $r^{*}$ & max $r^{*}$\\
\hline
6.06e-07 & 6.80e-07\\
6.34e-07 & 6.46e-07\\
5.51e-07 & 6.22e-07\\
5.87e-07 & 6.60e-07\\
5.10e-07 & 5.72e-07\\
4.71e-07 & 5.35e-07\\
5.10e-07 & 5.83e-07
\end{tabular}\\
\hline
\end{tabular}
}
\caption{Final global residual errors for $\varepsilon = 10^{-6}$ and $n = 150^{3}$.}
\label{tab:stab_r}
\end{center}
\end{table}
Table~\ref{tab:stab_wt} shows corresponding average execution times. We recall that, according to \cite{MagGBen2018}, the NFAIS2 provides slightly better execution times than the original algorithm from \cite{SavBert1996} which introduces few detection delays due to a first reduction operation prior to the proper snapshot phase (see, e.g., Algorithm~4 in \cite{MagGBen2018b}).
\begin{table}[!ht]
\begin{center}
{\small
\begin{tabular}{|c|c|c|c|}
\hline
& PFAIT & NFAIS2~\cite{MagGBen2018} & NFAIS5~\cite{MagGBen2018}\\
\hline
\begin{tabular}{c}
$p$\\
\hline
48\\
96\\
144\\
192\\
240\\
480\\
600
\end{tabular}
&
\begin{tabular}{c|c}
wtime (s) & $k_{\text{max}}$\\
\hline
46 & 15894\\
24 & 17353\\
16 & 17698\\
12 & 18122\\
10 & 17596\\
5 & 20356\\
4 & 19627
\end{tabular}
&
\begin{tabular}{c|c}
wtime (s) & $k_{\text{max}}$\\
\hline
61 & 16219\\
31 & 17445\\
21 & 17928\\
16 & 18256\\
13 & 18013\\
7 & 20504\\
6 & 20303
\end{tabular}
&
\begin{tabular}{c|c}
wtime (s) & $k_{\text{max}}$\\
\hline
61 & 16325\\
31 & 17533\\
21 & 18069\\
16 & 18282\\
13 & 17868\\
7 & 20826\\
5 & 19276
\end{tabular}\\
\hline
\end{tabular}
}
\caption{Wall-clock execution times for $\varepsilon = 10^{-6}$ and $n = 150^{3}$.}
\label{tab:stab_wt}
\end{center}
\end{table}
We can see here that
\[
\varepsilon - 0.2 \times 10^{-6} < r^{*} < \varepsilon + 1.9 \times 10^{-6}, \qquad \varepsilon = 10^{-6},
\]
which denotes a quite stable computational environment and suggests the possibility to reliably set an appropriate threshold $\varepsilon$ for ensuring Equation~\eqref{eq:ex:precision}. Such an investigation is additionally encouraged by the clearly better execution times obtained when applying no detection protocol.

Our next step was then to find the minimum threshold guaranteeing Equation~\eqref{eq:ex:precision}, which led us to
\[
\varepsilon = 4 \times 10^{-7},
\]
and observations in Table~\ref{tab:min_eps}.
\begin{table}[!ht]
\begin{center}
{\small
\begin{tabular}{|c|c|}
\hline
& PFAIT\\
\hline
\begin{tabular}{c}
$p$\\
\hline
48\\
96\\
144\\
192\\
240\\
480\\
600
\end{tabular}
&
\begin{tabular}{c|c|c|c}
min $r^{*}$ & max $r^{*}$ & wtime (s) & $k_{\text{max}}$\\
\hline
3.88e-07 & 6.28e-07 & 51 & 17386\\
4.06e-07 & 4.64e-07 & 26 & 18973\\
3.83e-07 & 7.90e-07 & 17 & 19298\\
4.24e-07 & 8.16e-07 & 13 & 19696\\
4.05e-07 & 5.83e-07 & 10 & 19060\\
4.29e-07 & 9.46e-07 & 6 & 22259\\
4.16e-07 & 8.54e-07 & 5 & 21221
\end{tabular}\\
\hline
\end{tabular}
}
\caption{Final global residual errors for $\varepsilon = 4 \times 10^{-7}$ and $n = 150^{3}$.}
\label{tab:min_eps}
\end{center}
\end{table}
However, as one can see on this example, it unexpectedly turned out that thresholds of the form $\alpha \times 10^{-7}$ with $\alpha \ne 1$ led us to a much less stable behavior, as we now have
\[
\varepsilon - 0.2 \times 10^{-7} < r^{*} < \varepsilon + 5.5 \times 10^{-7}, \qquad \varepsilon = 4 \times 10^{-7},
\]
which even almost violates the desired precision $10^{-6}$.

At last, according to such results about the characteristics of both the platform and the solver, we addressed the large-size problem as shown in Table~\ref{tab:large_r} and Table~\ref{tab:large_wt}.
\begin{table}[!ht]
\begin{center}
{\small
\begin{tabular}{|c|c|c|c|}
\hline
& PFAIT & NFAIS2~\cite{MagGBen2018} & NFAIS5~\cite{MagGBen2018}\\
\hline
\begin{tabular}{c}
$p$\\
\hline
144\\
192\\
240\\
360\\
480\\
600
\end{tabular}
&
\begin{tabular}{c|c}
min $r^{*}$ & max $r^{*}$\\
\hline
1.14e-07 & 1.81e-07\\
1.10e-07 & 2.12e-07\\
1.01e-07 & 2.15e-07\\
1.12e-07 & 1.91e-07\\
1.38e-07 & 3.11e-07\\
1.51e-07 & 2.01e-07
\end{tabular}
&
\begin{tabular}{c|c}
min $r^{*}$ & max $r^{*}$\\
\hline
5.28e-07 & 6.20e-07\\
6.03e-07 & 6.10e-07\\
5.22e-07 & 5.62e-07\\
5.12e-07 & 5.49e-07\\
3.81e-07 & 5.49e-07\\
4.05e-07 & 4.98e-07
\end{tabular}
&
\begin{tabular}{c|c}
min $r^{*}$ & max $r^{*}$\\
\hline
5.30e-07 & 6.01e-07\\
6.47e-07 & 6.96e-07\\
5.60e-07 & 6.14e-07\\
5.66e-07 & 7.13e-07\\
5.04e-07 & 5.56e-07\\
5.01e-07 & 5.98e-07
\end{tabular}\\
\hline
\end{tabular}
\hfill\\
\hfill\\
$\varepsilon = 10^{-6}$ for NFAIS2 and NFAIS5.\\
$\varepsilon = 10^{-7}$ for PFAIT.
}
\caption{Final global residual errors for $n = 185^{3}$.}
\label{tab:large_r}
\end{center}
\end{table}
\begin{table}[!ht]
\begin{center}
{\small
\begin{tabular}{|c|c|c|c|}
\hline
& PFAIT & NFAIS2~\cite{MagGBen2018} & NFAIS5~\cite{MagGBen2018}\\
\hline
\begin{tabular}{c}
$p$\\
\hline
144\\
192\\
240\\
360\\
480\\
600
\end{tabular}
&
\begin{tabular}{c|c}
wtime (s) & $k_{\text{max}}$\\
\hline
411 & 229848\\
315 & 243487\\
253 & 246580\\
168 & 240124\\
135 & 272549\\
114 & 293166
\end{tabular}
&
\begin{tabular}{c|c}
wtime (s) & $k_{\text{max}}$\\
\hline
437 & 186790\\
337 & 199160\\
275 & 203892\\
184 & 198632\\
147 & 225131\\
123 & 240475
\end{tabular}
&
\begin{tabular}{c|c}
wtime (s) & $k_{\text{max}}$\\
\hline
441 & 189613\\
338 & 199594\\
274 & 202271\\
183 & 199297\\
149 & 228574\\
124 & 243506
\end{tabular}\\
\hline
\end{tabular}
\hfill\\
\hfill\\
$\varepsilon = 10^{-6}$ for NFAIS2 and NFAIS5.\\
$\varepsilon = 10^{-7}$ for PFAIT.
}
\caption{Wall-clock execution times for $n = 185^{3}$.}
\label{tab:large_wt}
\end{center}
\end{table}
To therefore be on the safe side, we set the residual error threshold at $10^{-7}$ when using only successive non-blocking reduction operations. As a drawback, time saving was no longer maximized but the PFAIT still provided clearly better efficiency than snapshot-based termination protocols.

\section{Conclusion}
\label{sec:con}

The asynchronous convergence detection problem has been mainly investigated under worst-case theoretical aspects, leading to several more or less elaborated protocols. Early mathematical investigations led to applying few control on the communication pattern of the parallel iterations themselves. Less intrusive algorithmic approaches followed, however, most of them were either only optimistic or still intrusive at implementation level. To the best of our knowledge, the sole completely non-intrusive, distributed, approach always guaranteeing an accurate asynchronous convergence detection is the snapshot-based one.

Simple efficient snapshot-based protocols have recently been devised and implemented in an asynchronous iterations programming library. By analyzing such protocols, we showed here that, in practice, no detection protocol could be actually needed at all and that, instead, only non-blocking reduction operations could be successively applied as in classical synchronous computation. Indeed, it turns out that single-site computing centers are very likely to provide quite stable and reliable computational environments, even when dealing with node failures, which actually alleviates those worst-case abstract issues. In such cases then, it is interesting to see that, just as we experimented in this paper, asynchronous convergence detection can be simply achieved by turning the usual collective reduction operation into its non-blocking version, and that this leads to non-negligible overall execution time saving.

\section*{Acknowledgments}
The paper has been prepared with the support of the ``RUDN University Program 5-100'', the French national program LEFE/INSU, the project ADOM (M\'ethodes de d\'ecomposition de domaine asynchrones) of the French National Research Agency (ANR).

\bibliography{ref}
\bibliographystyle{abbrv}

\end{document}